\documentclass[12pt]{iopart}

\usepackage{url}
\usepackage{caption}
\usepackage{subcaption}
\usepackage[table]{xcolor}
\usepackage{tabularx}
\usepackage{mhchem} 
\usepackage{microtype}
\usepackage{hyperref}
\usepackage{graphicx}

\usepackage{siunitx}
\DeclareSIUnit\year{yr}
\DeclareSIUnit\photon{ph}
\DeclareSIUnit\electron{e}
\DeclareSIUnit\kev{\kilo\electronvolt}
\DeclareSIUnit\mm{\milli\meter}
\DeclareSIUnit\cm{\centi\meter}

\newcommand{\sit}[2]{$\SI{#1}{#2}$}
\newcommand{\gam}{$\gamma$}
\newcommand{\natt}{$^{22}$Na }

\newlength{\defwidth}
\begin{document}

\title[High-efficiency position resolved gamma ray detectors for 2D-ACAR]{High-efficiency position resolved gamma ray detectors for 2D-measurements of the angular correlation of annihilation radiation}

\author[1]{Kilian Brenner, Francesco Guatieri and Christoph Hugenschmidt}

\address{Heinz Maier-Leibnitz Zentrum (MLZ), Technical University of Munich, 	Lichtenbergstr. 1, 85748, Garching, Germany}

\eads{\mailto{Kilian.Brenner@frm2.tum.de}, \mailto{Francesco.Guatieri@frm2.tum.de}, \mailto{christoph.hugenschmidt@frm2.tum.de}}

\vspace{10pt}

\begin{abstract}
    The measurement of the 2D-Angular Correlation of Electron Positron Annihilation Radiation (ACAR) provides unique information about the bulk electronic structure of single crystals. 
    We set up a new prototype for 2D-ACAR measurements using two $24 \times 24$ ($\SI{26.8}{\mm} \times \SI{26.8}{\mm}$) pixelated LYSO scintillation crystals in combination with a glass light guide and $8 \times 8$ ($\SI{24}{\mm} \times \SI{24}{\mm}$) Multi Pixel Photon Counters (MPPCs).
    Compared to conventional Anger-cameras, typically comprising large \ce{NaI(Tl)} scintillators read out with photomultiplier arrays a larger implementation of our prototype would drastically improve resolution and count rate by taking advantage of the small pixel size of the scintillator, its much higher attenuation coefficient for \sit{511}{\kev} \gam-quanta and faster digital readout.
    With our prototype we achieved a detection efficiency of \sit{45}{\percent}, i.e. five times higher compared to NaI(Tl) used in our Anger cameras, leading to a 25 (!) times higher coincidence count rate in ACAR measurements.
    A spatial resolution of \sit{1}{\mm} was obtained, which is limited by the pixel size of the scintillator. 
    We demonstrate the high performance of the setup by (i) imaging the local distribution of \natt in a proton-irradiated aluminum target and (ii) determining the Fermi energy of Cu from 2D-ACAR spectra recorded for a polycrystalline copper sample.

\end{abstract}

\vspace{2pc}

\maketitle

\section{Introduction} \label{sec:introduction}
    Knowledge of the electronic structure of solids, in particular of the Fermi surface (FS), is essential for a microscopic understanding of electronic correlations, which lead, e.g., to magnetic and structural phase transitions in Heusler alloys \cite{santhoff1995}, the development of magnetic order and/or superconductivity \cite{pfleiderer2009} and the development of novel electronic states in magnetic systems \cite{lohneysen2007}. These transitions are often accompanied by dramatic changes of the transport properties. 
    The Fermi surface can be determined by several well established techniques such as de Haas-van Alphen (dHvA) measurements \cite{shoenberg2007}, Compton scattering \cite{cooper1985, ketels2021}, ARPES \cite{damascelli2003} and ACAR \cite{hautojarvi1979, ceeh2016, weber2015, ketels2021physStat}. Each technique has its merits and drawbacks, which have to be taken into account. For example, dHvA requires low temperatures, high magnetic fields, or long mean free paths of the electrons, i.e., very clean materials; Compton scattering delivers a 1D projection of the electron momentum distribution making long measuring times necessary for the determination of the full FS; ARPES is extremely sensitive to surface contamination and difficult to apply in the bulk. ACAR has the main advantage that special conditions such as high magnetic fields, low temperatures and long mean free paths of the electrons are not required. Its drawback, however, is the need of high-intensity positron sources or beams.
    
    2D-ACAR is applied to determine the bulk electronic structure by using positrons. After implantation into the sample, the positrons thermalize rapidly within a few picoseconds and subsequently propagate in Bloch states, as they also feel the crystal potential. The annihilation of these thermalized positrons with electrons leads predominately to the emission of two \sit{511}{\kev} photons. In the center-of-mass (rest) frame of the electron-positron pair, the photons are emitted in opposite directions with equal energy due to conservation of momentum. However, in the laboratory frame, the transverse (i.e. perpendicular to the emission direction of the photons) components of the electron and positron momenta lead to a proportional deviation of the emission angle from \sit{180}{\degree}. Hereby, the small angle approximation is valid because the electron and positron momenta are small in comparison with the momentum of a \sit{511}{\kev} photon.
    
    For 2D-ACAR measurements, both annihilation quanta are detected simultaneously with two spatially resolved detectors (so-called Anger cameras). A single ACAR-spectrum hence delivers a planar 2D-projection of the electron momentum distribution. The 3D distribution of the electron momenta in the sample can be reconstructed by spectra recorded at several angles of the sample with respect to the detector-detector axis. However, the measurement times of conventional 2D-ACAR experiments is extremely long (several days per spectrum) due to the needed high momentum resolution achieved with at a large baseline of the detectors ($\SIrange{15}{20}{\m}$, see e.g., \cite{ceeh2013, weber2013}).    
    
    In this paper, we present an approach that allows us to overcome the extremely long measurement times by implementing a new position-resolving detection system with a spatial resolution of \sit{1}{\mm}. 
    Moreover, we achieved a $25$ fold increase in detection efficiency compared to a pair of conventional Anger-cameras leading to accordingly faster measurements.

\section{Design and Setup of the Detectors}

\subsection{The Scintillation Detectors} \label{sec:detStructure}

\begin{figure}
        \centering
        \includegraphics[width=0.9\defwidth]{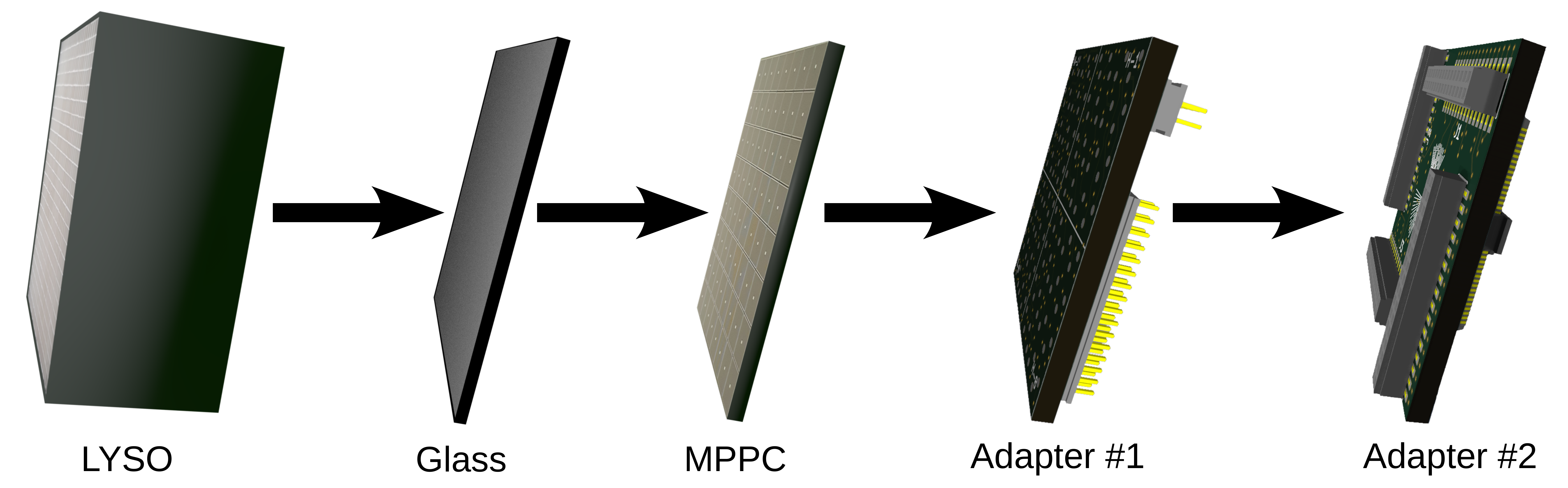}
        \caption{Complete detector stack of a single detector: From left to right: pixelated LYSO scintillator ($\SI{24}{\mm} \times \SI{24}{\mm} \times \SI{20}{\mm}$), glass plate (\sit{2}{\mm}), energetically and spatially resolved MPPC, two adapters to interface the TOFPET2 readout.}
        \label{fig:detstack}
    \end{figure}

    The whole detector comprises a LYSO scintillator, a thin glass light guide, a Multi-Pixel-Photon-Counter (MPPC) and two adapters which have been custom designed for this setup to allow for easier debugging (see figure \ref{fig:detstack}). The scintillator, the light guide and the MPPC are connected with optical grease. The rightmost adapter in figure \ref{fig:detstack} connects the detector to the TOFPET2 readout by PETsys electronics. As MPPCs are extremely light sensitive, the entire detector stack is housed in a 3D-printed light shield.
    
    Given their deployment in clinical positron-emission tomography (PET) applications, Anger-cameras have undergone a rapid evolution. The state-of-the-art scintillation material is lutetium-yttrium oxyorthosilicate (LYSO) which exhibits superior physical properties compared to other scintillators: (i) an excellent absorption coefficient required for high spatial resolution and efficiency, (ii) high peak-to-total ratio, i.e. high photo-peak efficiency that results in a high count rate in the \sit{511}{\kev} photo peak, and (iii) high light yield that results in a good energy resolution and allows the distinction of Compton-scattered gamma quanta. A comprehensive list of the crystal properties as well as a comparison with the commonly used \ce{NaI(Tl)} scintillator is shown in table \ref{tbl:scintillators}.
    Furthermore, thick pixelated LYSO crystals with a reflective layer between the small crystals are commercially available.

    \begin{table}
	\begin{center}
	\begin{tabular}{c|c|c}
		& LYSO & \ce{NaI} \\
		\hline
		Density [\sit{}{\gram\per\cubic\cm}] & 7.25 & 3.67 \\
		Wavelen. of emission peak [\sit{}{\nano\meter}] & 420 & 415 \\
		Light yield [\sit{}{\photon\per\MeV}] & \sit{29000}{} & \sit{40000}{} \\
		Decay time [\sit{}{\nano\second}] & \sit{42}{} & \sit{264}{} \\
		Refractive index & \sit{1.82}{} & \sit{1.85}{} \\
		Tot. att. coeff. (at \sit{511}{\kev}) [\sit{}{\per\cm}] & \sit{0.80}{} & \sit{0.33}{} \\
		Photo-effect frac. (at \sit{511}{\kev}) [\sit{}{\percent}] & \sit{33.8}{} & \sit{18.2}{} \\
		Energy res. (at \sit{662}{\kev}) [\sit{}{\percent}] & \sit{10.9}{} & \sit{7.2}{} \\
	\end{tabular}
	\end{center}
	\caption{Important characteristics of LYSO and \ce{NaI} scintillation crystals. Values taken from \cite{NIST_gam, LYSO_datasheet, NaI_datasheet}. \label{tbl:scintillators}}
	\end{table}

    To measure the number of optical photons generated within the scintillator, an array of silicon photomultipliers (SiPMs) -- also called MPPCs -- is employed. Each of the MPPC elements is capable of resolving energy, allowing for a center-of-gravity (COG) calculation that in turn enables a spatial resolution down to the LYSO crystal pixel size. These MPPCs offer high readout speed and time resolution, while exhibiting a compact and cost-effective design.
    Each MPPC element consists of several thousand Avalanche-Photo-Diodes (APDs) connected in series \cite{S13361}. The APDs are biased in reverse with an overvoltage (OV) higher than their breakdown voltage. A single photon can create an electron-hole pair inside the APD, causing it to become conductive until the resulting electron avalanche is stopped by quenching resistors \cite{mppc_kapd9005e}. Measuring the charge generated by all APDs combined, allows to gives a measure of the number of photons hitting the detector and thereby provides the energy resolution of the MPPC.
      
    Each of the two detectors comprises a single pixelated LYSO crystal with the dimensions of $\SI{26}{\mm} \times \SI{26}{\mm} \times \SI{20}{\mm}$ and $24 \times 24$ pixels.
    Crystal pixels are separated by a \ce{BaSO4} reflector which acts as a diffuse scatterer. A \sit{2}{\mm} thick glass pane installed between the crystal and the MPPC detector enables photons that are created in one of the crystal pixels to spread across multiple detector elements to allow for a COG calculation. These photons are then detected by the MPPC detector (Hamamatsu S13361-3050AE-08) with dimensions of $\SI{24}{\mm} \times \SI{24}{\mm}$ ($8 \times 8$ elements). The digitization of the MPPCs is conducted with hardware from PETSys electronics, which was originally developed for clinical PET applications and offers an intrinsic time resolution of \sit{30}{\ps} \cite{PETsys_asic}.

\subsection{Measurement Setup}
\begin{figure}
        \centering
        \includegraphics[width=0.9\defwidth]{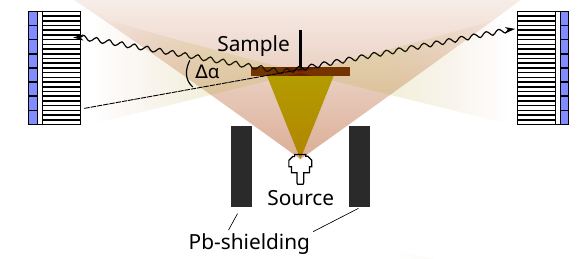}
        \caption{Simple setup to demonstrate first 2D-ACAR measurements.
        The lead shielding blocks the radiation from the \natt source without affecting the annihilation \gam-quanta emitted from the sample. The angle deviation from \sit{180}{\degree} is measured by the angles $\Delta\alpha$ and $\Delta\beta$ ($\Delta\beta$ is perpendicular to the plane of $\Delta\alpha$, not shown).} 
        \label{fig:setupAcar}
    \end{figure}

    Measurements for the detector characterization were conducted by illuminating the detectors homogeneously with a \natt source.
    For the 2D-ACAR measurement, two of the detectors described in section \ref{sec:detStructure} were placed at a distance of \sit{87}{\cm} to the copper sample. As positron source, the same \sit{60}{\mega\becquerel} encapsulated \natt source was used. Additional lead bricks were placed to shield any radiation stemming from annihilations inside the \natt source itself. A schematic of the setup is shown in figure \ref{fig:setupAcar}.
 
   \begin{figure}
    \centering
    \includegraphics[width=0.7\defwidth]{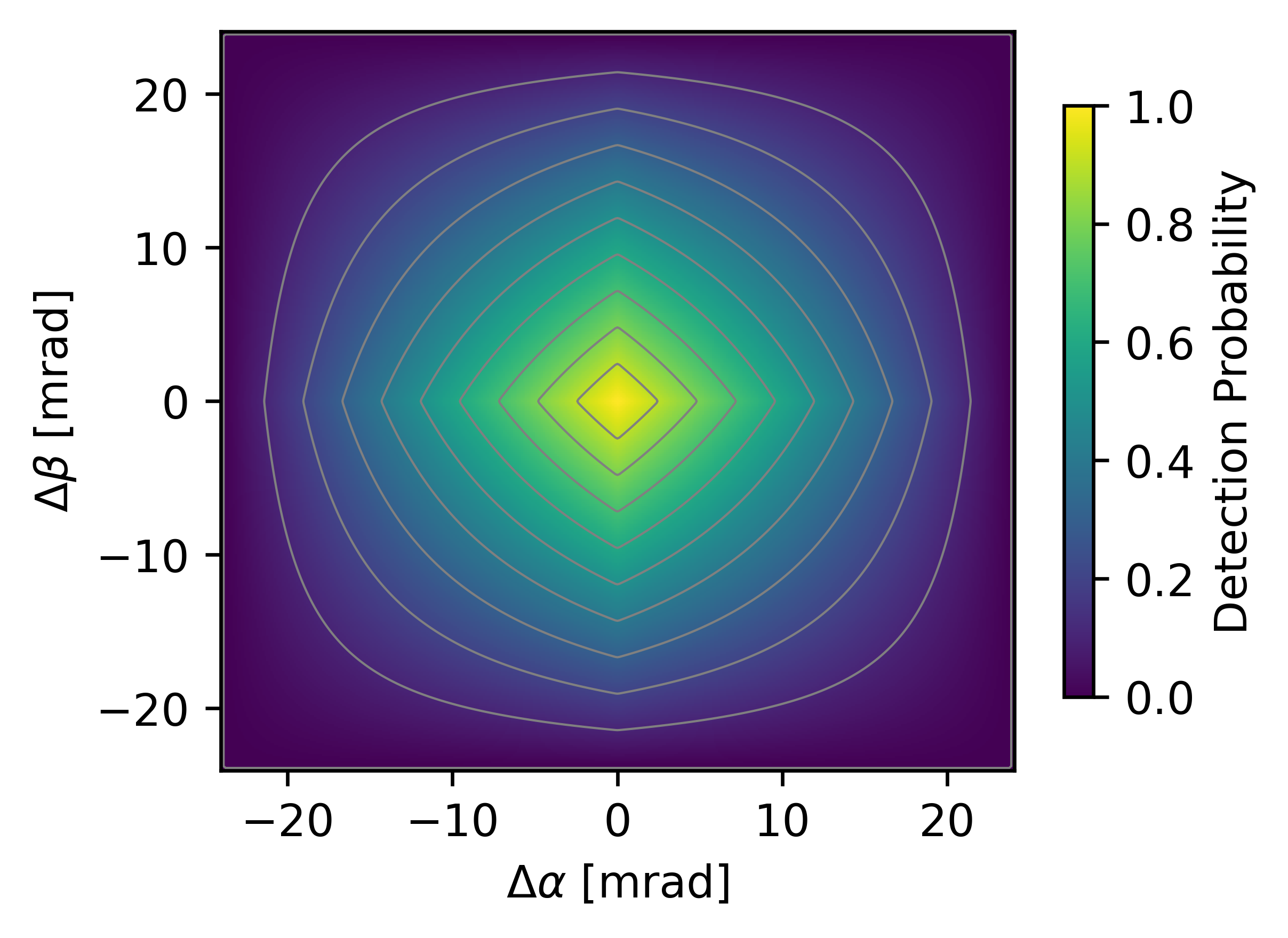}
    \caption{Momentum sampling function (MSF) describing the combined detection probability of the two detectors at a sample-detector distance of \sit{1}{\meter}. The angles $\Delta\alpha$ and $\Delta\beta$ are the difference from \sit{180}{\degree} between the two detected \gam-quanta.
    The MSF shown is originated purely from geometric considerations, i.e. assuming a homogeneous detector response.}
    \label{fig:msf}
\end{figure}

    The angle between the two emerging gamma quanta from a positron-electron annihilation event is determined by measuring their positions with Anger cameras in coincidence.    
    Due to the finite detector size, not all angles are equally likely to be recorded. In ACAR measurements, this effect has to be accounted for by dividing the recorded spectra by the so called Momentum-Sampling-Function (MSF). Given the detector efficiencies as a function of the angles $\epsilon_1(\alpha_1, \beta_1)$ and $\epsilon_2(\alpha_2, \beta_2)$ and using small-angle approximation $\alpha \approx \Delta x_1-\Delta x_2$ and $\beta \approx \Delta y_1 - \Delta y_2$, the MSf becomes the 2-dimensional convolution of both detector efficiencies:
	
	\begin{equation}
		MSF(\alpha, \beta) = \int \int \epsilon_1(\alpha - \alpha', \beta - \beta') \cdot \epsilon_2(\alpha', \beta') \, d\alpha' \, d\beta'
	\end{equation}
    
    A purely geometric MSF (i.e. neglecting detector imperfections) at a detector-sample distance of \sit{1}{\m} is shown in figure \ref{fig:msf}.

\section{Characteristics and Performance of the Detectors}

    In the following measurements, the detector was uniformly illuminated by a \sit{511}{\kev} \gam-source positioned far enough away so that the angles of incidence can be assumed to be orthogonal to the detector surface. As both detectors are identical in construction and performance, we will only show measurements from one of the two detectors.

\subsection{Spatial Resolution}

\begin{figure}
        \centering
        \includegraphics[width=0.9\defwidth]{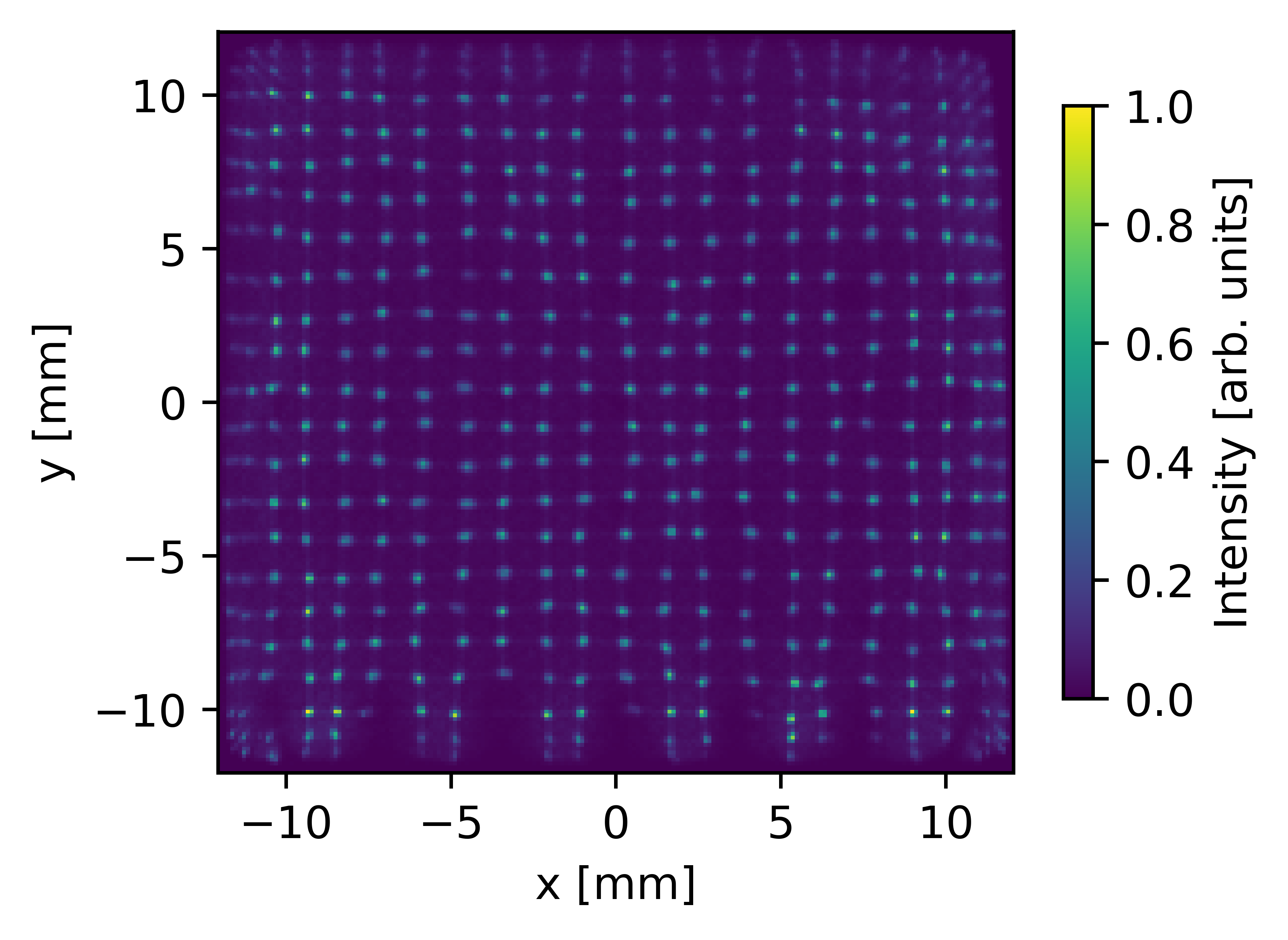}
        \caption{Histogram of reconstructed incident \gam\: positions of a homogeneously illuminated detector. Due to the pixelated nature of the LYSO crystal, the reconstructed positions tend to center in their respective pixels, forming the visible spots, mostly caused by the difference in their energy sensitivity.}
        \label{fig:HomIll}
    \end{figure}

    The \gam\: interaction points inside the LYSO scintillator were reconstructed for a homogeneous detector illumination using a COG calculation. The resulting histogram for one of the two detectors is shown in figure \ref{fig:HomIll}.

    A grid of distinctive points is visible, where each point corresponds to a single crystal pixel. These points will be called event groups in the following. The existence of these event groups is indicative but not a sufficient criterion to establish a resolution limit of the crystal pixel size (\sit{1}{\mm}). Even though the diffuse reflector (\ce{BaSO4}) should make any spatial distinction inside a single crystal pixel impossible, we conducted another experiment to show that.

    With the following measurement we wanted to observe whether the photon output of a \sit{1}{\milli\meter} $\times$ \sit{1}{\milli\meter} crystal pixel is indeed independent of the incident \gam\: position within the crystal pixel.
	For this purpose, a \sit{5}{\cm} thick lead brick approximately \sit{70}{\centi\meter} away from the \sit{60}{\mega\becquerel} source was placed directly in front of a detector to partially block radiation from hitting a single crystal pixel. In consecutive measurements, the lead brick was moved with a stepper motor in increments of \sit{100}{\micro\meter} over the range of \sit{3}{\mm}.

	The lead brick was moved back and forth for $30$ iterations in total and the results were accumulated before further processing.
    From this measurement we extracted the data only corresponding to one of the pixels that was partially blocked by the lead brick.

\begin{figure}
        \centering
        \includegraphics[width=0.8\defwidth]{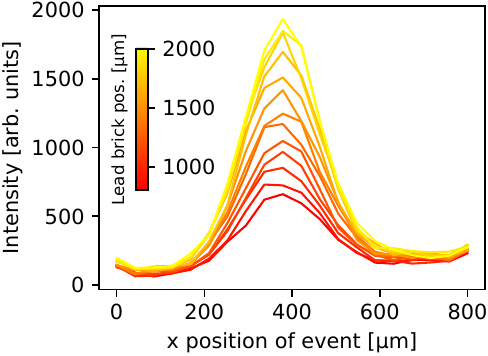}
        \caption{A lead brick was moved in steps of \sit{100}{\micro\meter} in front of a crystal pixel from a position blocking all radiation from the crystal pixel (red) to a fully exposed crystal pixel (yellow). Histograms of the reconstructed x-position for an event group of the respective crystal pixel clearly show that the shape of the spectrum does not depend on the incident \gam\: position.}
        \label{fig:pos_dist_proj}
    \end{figure}

    Figure \ref{fig:pos_dist_proj} shows the distribution of reconstructed positions of the \gam-quanta inside an event group corresponding to the crystal pixel that was partially exposed to radiation, projected onto the x-axis to increase visibility.
    The histograms color encodes the lead-brick position from a fully exposed crystal pixel (yellow) to a not illuminated crystal pixel (red).
    It is clearly visible that the shape of the histogram does not change depending on which part of the crystal pixel was exposed to radiation, i.e. there is no dependence on the position of the incident \gam-quanta absorbed inside the pixel. Even the fully covered crystal pixel shows a weak signal, which is caused by (Compton-) scattered \gam-quanta from neighboring crystal pixels.
    This leads to the conclusion that no spatial information below the crystal pixel size can be extracted. The spatial resolution is hence tied to the crystal pixel size (in our case: \sit{1}{\mm}).
        
\subsection{Energy Resolution}

    \begin{figure}
        \centering
        \includegraphics[width=0.99\defwidth]{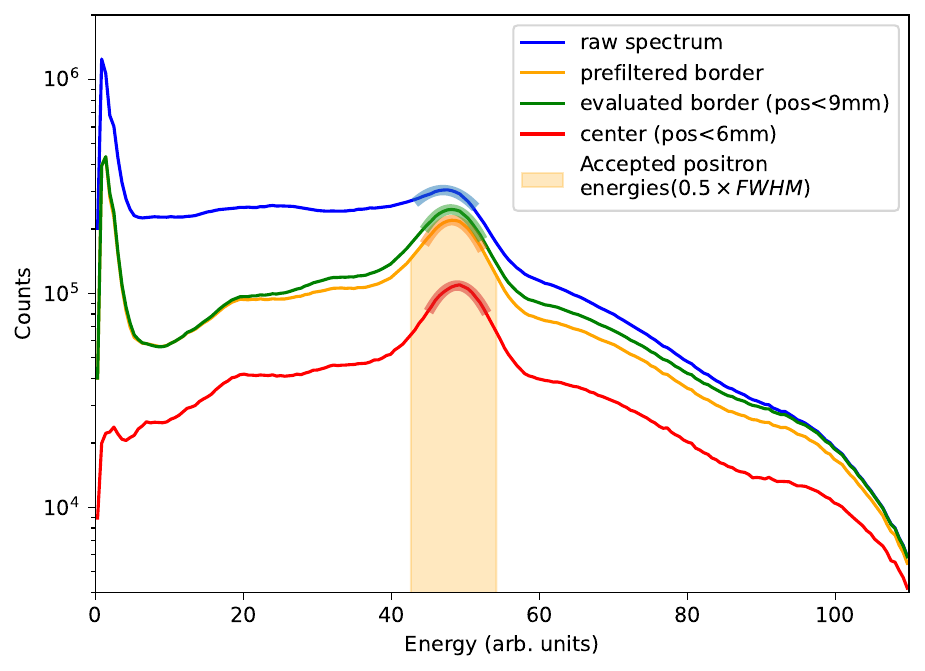}
        \caption{Distribution of the collected energy of a homogeneously illuminated detector. The blue curve is the recorded raw spectrum. The data can be filtered by successively removing events too close to the edge of the detector (green, orange and red curve). To determine the energy range of accepted events, the prefiltered spectrum is fitted with a Gaussian distribution and a fraction of the FWHM is used for energy selection (orange area). A Gaussian fit of the \sit{511}{\kev} peak within the fit region is shown as thicker line in all curves.}
        \label{fig:energyRes}
    \end{figure}

    The detector was again uniformly illuminated from the same \sit{60}{\mega\becquerel} \natt source positioned at a distance of around \sit{80}{\cm}. 
    The number of \sit{511}{\kev} \gam-quanta hitting the detector was about $\SI{4.3e3}{\per\second}$.

    The energy spectrum is created by summing the energies of each element associated to the same event. This -- we will call it raw energy spectrum -- was recorded for \sit{40}{\minute} and is shown in figure \ref{fig:energyRes} as a blue curve.
    When a \gam-quantum impinges close to the detector border, a portion of the optical photons generated in the LYSO crystal will not reach any element of the MPPC and therefore distort the energy histogram.
	To produce a less biased spectrum, events can be filtered based on their position before creating the spectrum. The green curve in figure \ref{fig:energyRes} contains only events with a reconstructed position at least \sit{3}{\mm} (i.e. the size of one MPPC element) apart from the MPPC border. The energy histogram of those events whose coordinates are more than \sit{6}{\milli\meter} apart from the crystal border is displayed in the red curve.
	The third curve shown in figure \ref{fig:energyRes} is obtained by a so-called pre-filter. It does not discriminate events based on their reconstructed positions but rather excludes all events whose element with the highest recorded energy is at the border of the MPPC. In all following spectra, the pre-filter is used (unless otherwise stated), since it yields a histogram with much less distortion from the expected \natt spectrum, while also being the most simple filtering method.
	
	As expected, filtering affects the low-energy portion of the spectrum much more than the high-energy one. The pre-filtered histogram and the "evaluated border" (see figure \ref{fig:energyRes}) are quite similar and both of them reduce a significant fraction of dark counts and are able to reproduce the \sit{511}{\kev} peak better. The strongest filter, which allows only for even more centered events, removes most of the low-energy events. This indicates that the low-energy peak showing in the other curves is comprised mostly of events impinging on the edge of the detector and does not hold any physical significance.
	
	The peaks of all four spectra were compared by fitting them with a Gaussian distribution. The best fit for all curves is shown in figure \ref{fig:energyRes} as a thicker line; plotted only within the horizontal range of data used to perform the fit itself. The unfiltered spectrum shows a FWHM of $\approx 36\%$, but it is apparent that the Gaussian model is not very suitable to fit this data. For all three other spectra the FWHM is $\approx 25\%$ with only minor differences. This represents the actual energy resolution of the whole detector.
	
	An energy spectrum is created and fitted for each measurement to determine the energy range for positrons before proceeding with the data analysis. The most convenient method is to apply the pre-filter to the data and perform a Gaussian fit. The resulting energy range (here: $0.5 \times \text{FWHM}$) is indicated in figure \ref{fig:energyRes} as an area highlighted in orange.

    This procedure is performed for all measurements to exclude changes in the environment that can affect the position of the peak. For this purpose, an automated algorithm which uses the data from the respective measurement is applied to calibrate the detectors before further data processing.

\subsection{Temperature Dependence}

    The gain, i.e. the charge generated per photon, of MPPCs depends strongly on their temperature. To quantify this effect, energy histograms were recorded at different sensor temperatures. 
    For simplicity, the detectors were heated to different temperatures to run these measurements.
    After changing the temperature setting of the heating control system, the setup was given multiple hours to stabilize before a measurement. 
    The temperatures values are recorded by the TOFPET2 system which employs temperature sensors at the connector close to the MPPC.

    \begin{figure}
        \centering
        \includegraphics[width=0.9\defwidth]{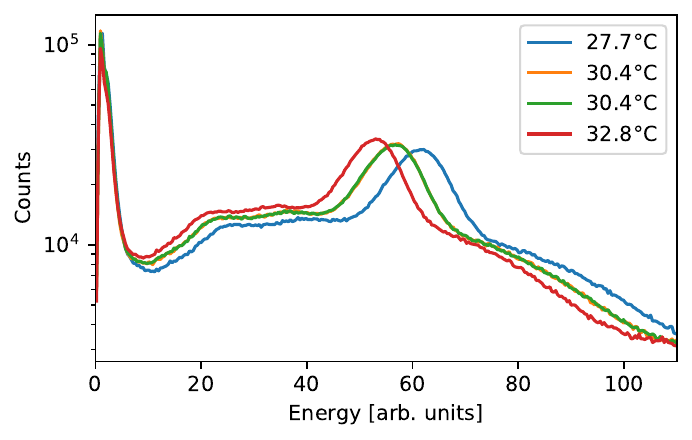}
        \caption{Effects of sensor temperature on the energy histogram. Higher temperatures apparently shift the spectrum towards lower energies, because higher temperatures increase the breakdown voltage of an APD leading to an effective reduction of the overvoltage (OV) \cite{Schug_2016}. The temperature of \sit{30.4}{\degreeCelsius} was measured twice with a time difference of \sit{13}{\hour} to show the stability of the measurement (orange curve coincides with green one.}
        \label{fig:tempDep}
    \end{figure}

    The temperature dependent energy histograms are shown in figure \ref{fig:tempDep}. Measurement time for each detector was \sit{10}{\minute}.
    The sensor temperature ranges between \sit{27.7}{\degreeCelsius} and \sit{32.8}{\degreeCelsius}. The two measurements at a temperature of \sit{30.4}{\degreeCelsius} were performed \sit{13}{\hour} apart to evaluate reproducibility; their spectra match perfectly. This is a strong indicator that the temperature is the only significant ambient effect for measurements, which is to be expected.
	
	Figure \ref{fig:tempDep} shows clearly that the spectrum gets more compressed for higher temperatures. This can be explained by the fact that the breakdown voltage of the APDs in the MPPC increases with a higher temperature. This was shown e.g. in \cite{Schug_2016} for a similar detector. Increasing the sensor-temperature therefore effectively lowers the overvoltage (OV) applied to the SiPM which reduces the gain and hence compresses the spectrum towards lower energies.	
	Especially for longer measurements -- where the temperature might not be constant -- this effect has to be accounted for. For the future bigger detector a cooling system is planned to be employed. This can also significantly reduce dark counts \cite{piatek2014physics}.

\subsection{Background signal}

    \begin{figure}
        \centering
        \includegraphics[width=0.9\defwidth]{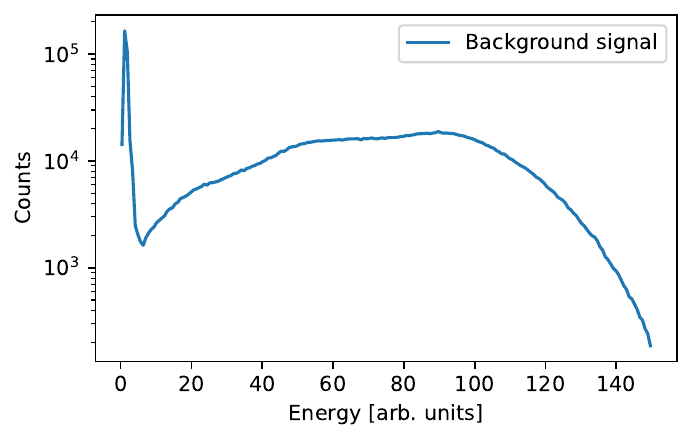}
        \caption{Energy histogram of a background measurement of the detector.}
        \label{fig:background}
    \end{figure}

    Due to their high sensitivity, MPPCs are also very sensitive to dark counts. 
    In order to quantify the influence of the background, we performed multiple background measurements as exemplarily shown in figure \ref{fig:background}. This measurement was not prefiltered. The measurement time was \sit{10}{\minute} and the MPPC temperature was around \sit{40}{\degreeCelsius}.
    Most background events have very low energy, which, however, are not problematic, because they can easily be filtered.

    In order to exclude light bleeding through the casing and background radiation as possible causes of this background, we conducted the same measurement with an additional lead shielding layer as well as another light shielding layer around the one already in place. Both of these modifications did not result in a change of the background signal.
    The most likely cause of the signal are dark counts. Dark counts in a pixel/element have a certain probability of cross-talk with other pixels/elements, resulting in a pile up of dark-counts and hence leading to a background with higher energies.
    Elements are grouped as an event, when they are firing within a defined time window. Lowering this time window did also not change the signal significantly.

\section{Measurements and Results}
    To showcase the capabilities of the detector prototype and the feasibility for an actual detector in the future, two benchmark measurements were conducted. 
    Due to the limited detector size and a low detector-source distance, however, the resolution is inherently limited.
    
    In the first one we evaluate the spatial distribution of \natt produced within a thin aluminum plate by proton-beam irradiation \cite{krug2024}. Rather than measuring the angle between the two coincident \gam-quanta, a \sit{180}{\degree} angle is assumed to calculate the origin position of the positron emitted from \natt and annihilated inside the sample.

    The second measurement is an actual 2D-ACAR measurement on polycrystalline copper at room temperature to demonstrate the reliable determination of the Fermi energy. Here the angle between the two \gam-quanta is the variable of interest. For that, the origin position of the positron has to be known.
    The measured angle and the transverse electron momentum $p_\bot$ have the following dependency:

    \begin{equation} \label{eq:momentum}
        \Delta\alpha = \arctan \left( \frac{p_\bot}{m_0 c} \right) \approx \frac{p_\bot}{m_0 c}
    \end{equation}
    where $c$ is the speed of light and $m_0$ the electron rest mass.

\subsection{Distribution of $\,^{22}$Na in Proton Irradiated Aluminum}

    This measurement evaluated the spatial distribution of \natt within a \sit{0.3}{\mm} thin aluminum plate that was irradiated with a proton beam \cite{krug2024}. 
    Similar to PET measurements, the \natt distribution can be measured by computing the origin positions of the \gam-quanta by assuming a \sit{180}{\degree} angle between them. 
    The proton beam used to irradiate the sample is assumed to have an elliptical shape of around $\SI{3}{\mm} \times \SI{6}{\mm}$, which should produce a slightly larger \natt distribution according to simulations performed in \cite{krug2024}. The measured dimensions of the \natt area in the aluminum disk have a FWHM of \SI{4.13(22)}{\mm}$  \times  $\SI{7.50(09)}{\mm}  \cite{krug2024}.
         
\begin{figure}
        \centering
        \includegraphics[width=0.9\defwidth]{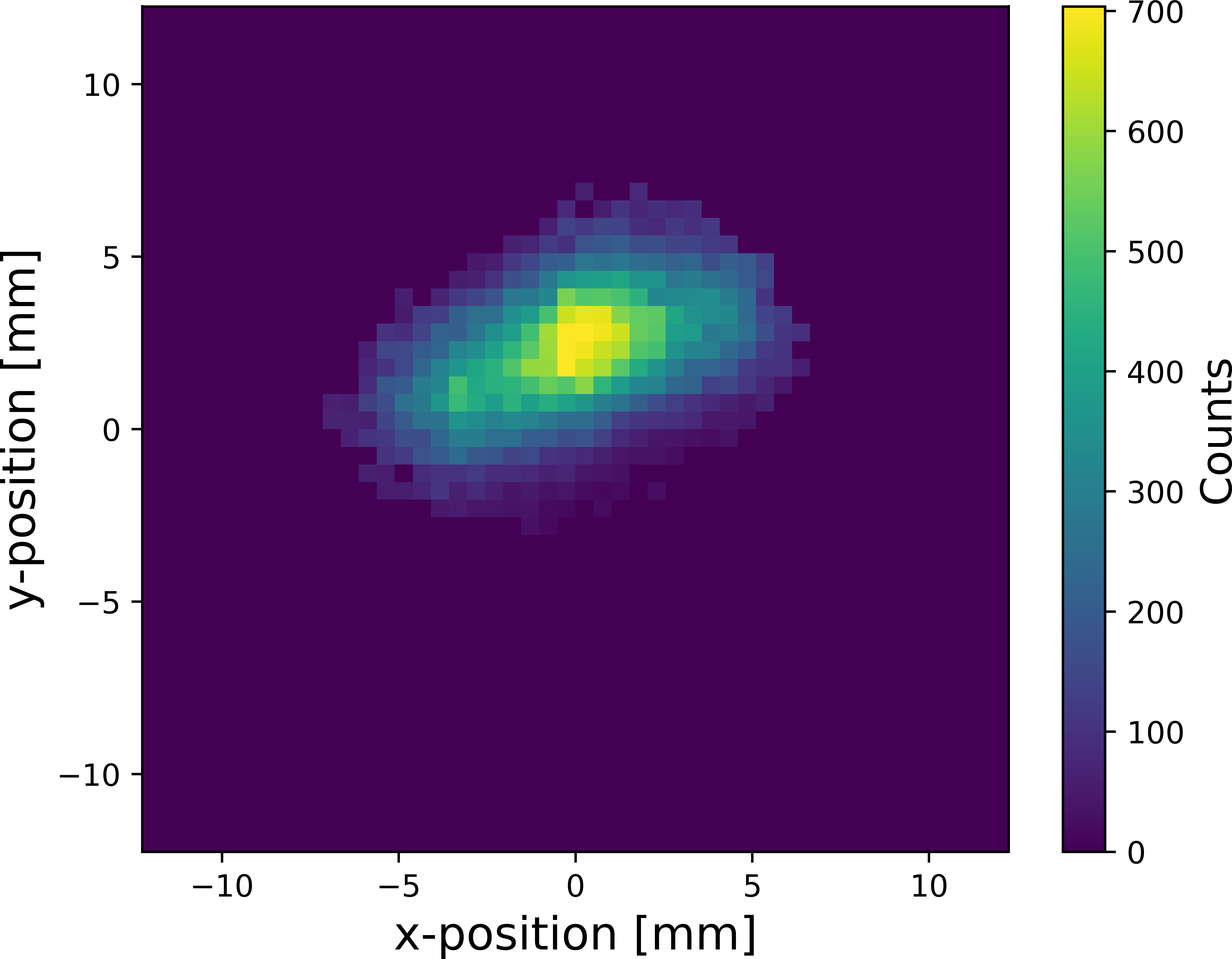}
        \caption{Histogram of reconstructed positions of the positron-electron annihilation. By assuming a perfect \sit{180}{\degree} angle between the emitted \gam-quanta, their annihilation position -- and with that the \natt distribution -- can be measured. The data was corrected by the position dependent detection efficiency. 
        A 2D Gaussian fit results in an elliptically shaped  \natt distribution of $\SI{4.2(3)}{\mm} \times \SI{7.7(4)}{\mm}$ (FWHM).}
        \label{fig:aluminum}
    \end{figure}

    In this measurement, the detectors were placed \sit{15}{\mm} apart from the sample -- which has an activity of approximately \sit{4.6}{\kilo\becquerel} -- and their signal was recorded for \sit{48}{\hour}.
    The reconstructed origin positions of the \gam-quanta are shown in figure \ref{fig:aluminum}.
    The shown data has already been corrected to account for the varying detection efficiency for different \gam\: origin positions (not to be confused with the MSF).

    By fitting a simple 2D Gaussian distribution, we obtain a \natt shape of $\SI{4.2(3)}{\mm} \times \SI{7.7(4)}{\mm}$ (FWHM). Both of these values are in excellent agreement with the measurements performed in \cite{krug2024}.

\subsection{2D ACAR of Polycrystalline Copper}

    \begin{figure}
        \centering
        \includegraphics[width=0.9\defwidth]{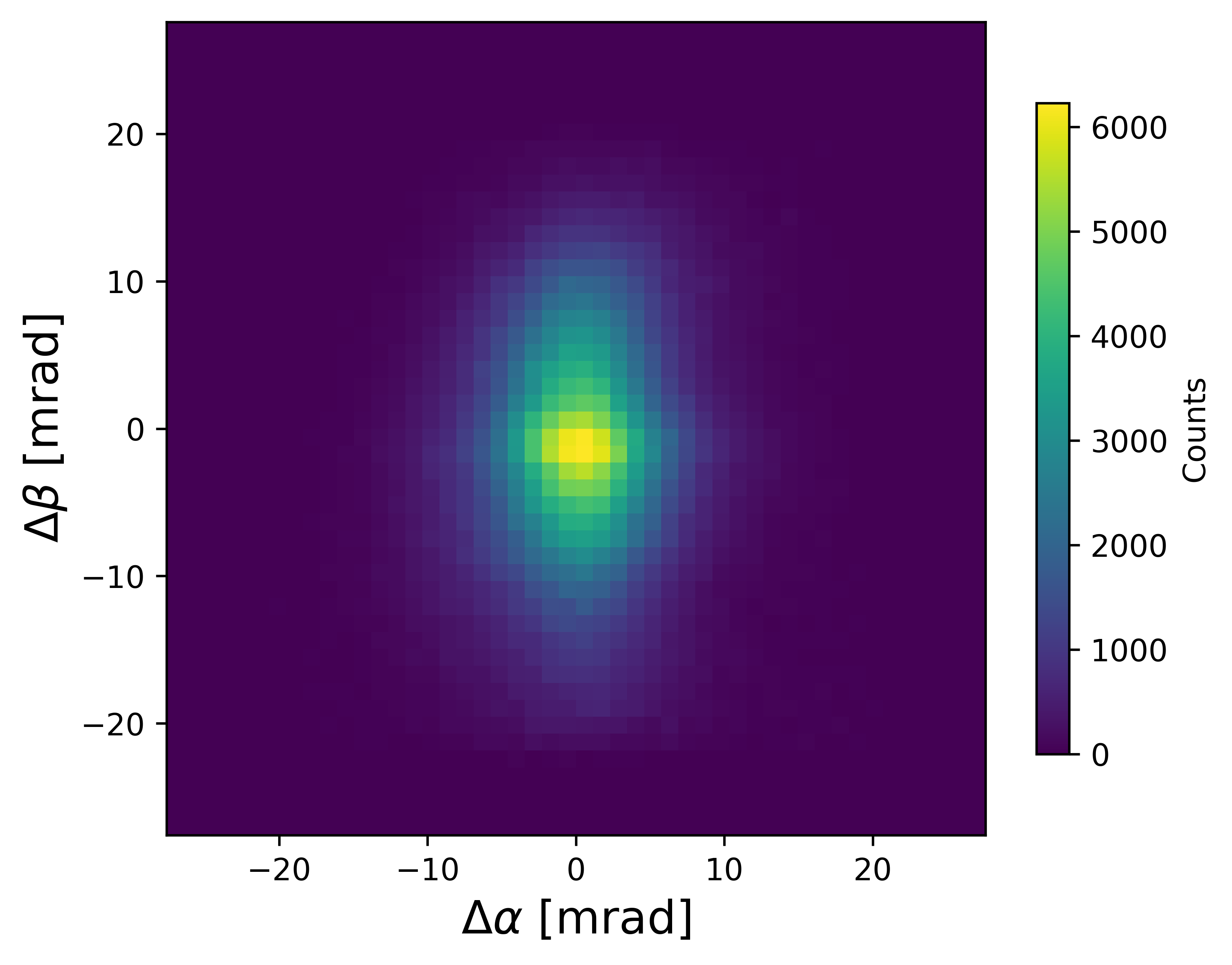}
        \caption{2D-ACAR spectrum of a polycrystalline copper sample at a distance of \sit{870}{\mm} to the sample. The spectrum is elongated in the direction of the $\beta$ angle due to the spatial extension of the area of the sample irradiated with positrons in this direction.}
        \label{fig:acar}
    \end{figure}
    
    As a proof of concept for future applications based on bigger detectors, we performed a 2D-ACAR experiment on a polycrystalline copper sample. Since all crystal directions are expected to be distributed evenly, the Fermi-sphere should be perfectly round, i.e without showing any anisotropic features. For the measurement we used a \natt point source inside a metallic capsule with a thin Ti window, from which positrons are emitted towards the copper sample. The distance between positron source and sample is a few \sit{}{\mm}.
    The \gam-quanta resulting from positrons annihilating inside the source capsule itself were shielded using lead bricks (see figure \ref{fig:setupAcar}). 

    The 2D-ACAR spectrum shown in figure \ref{fig:acar} was recorded for \sit{10}{\hour} at a detector-sample distance of \sit{87}{\cm} and is already corrected with the MSF. The angle distribution is elongated in the $\beta$ direction. This is caused by the spatial extension of the sample in this direction. Along the $\alpha$ angle, the emission region looks point-like as it only extends to the positron penetration depth in copper. Since the probed depth only amounts to a few ten \sit{}{\micro\meter} it can be neglected in the data evaluation.
    
\begin{figure}
        \centering
        \includegraphics[width=0.9\defwidth]{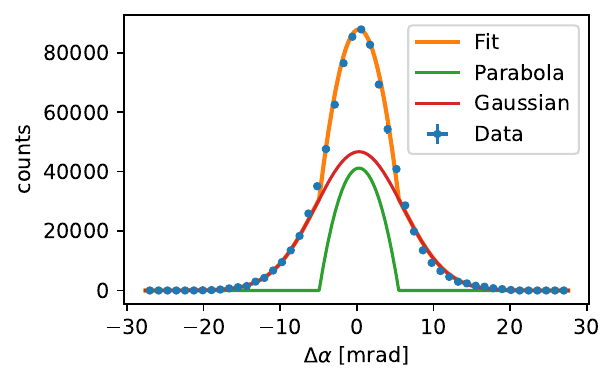}
        \caption{1D projection of the 2D-ACAR measurement of polycrystalline copper shown in figure \ref{fig:acar}. The data is fitted with the sum of a Gaussian background and a parabola.
        The resulting Fermi-energy given by the cut-off of the parabola at the Fermi momentum amounts to \sit{6.8(8)}{\electronvolt} in agreement with the literature value of \sit{7.0}{\electronvolt}.}
        \label{fig:1dacar}
    \end{figure}

    Figure \ref{fig:1dacar} shows the 1D projection of the 2D-ACAR spectrum over the $\alpha$ angle and hence eliminates the $\beta$ distrubition. This spectrum shows a parabola which originates from conduction electrons in the Fermi-sphere. Annihilations with bound electrons produce a Gaussian-shaped background. 

    The maximum angle of the fitted parabola is \sit{5.14(32)}{\milli\radian}, resulting in a Fermi-energy of \sit{6.8(8)}{\electronvolt}. This agrees with the value of \sit{7.0}{\electronvolt} found in literature \cite{kittel}.
        A more complex fit of the 2D-distribution that also accounts for the elongation caused by the positron beam width by assuming a rectangular spot, yields a Fermi-energy of \sit{6.9(4)}{\electronvolt}.
    
\section{Summary}
    We have set up a prototype for the coincident and spatially resolved detection of \sit{511}{\kev} annihilation radiation with high efficiency.
    Based on our detection scheme, future setups using larger detectors will enable 2D-ACAR measurements with considerably improved performance. 
    By using \sit{20}{\mm} thick LYSO crystals the detection efficiency for \sit{511}{\kev} gamma quanta in the photo peak is increased by a factor of five compared to \ce{NaI(Tl)} used in conventional Anger cameras. This would lead to a $25$ times (!) higher coincidence count rate in 2D-ACAR measurements. 
    In addition, using $1 \times 1$ \sit{}{\mm} LYSO crystals improves the position resolution by a factor of $3.5$.
    This either enhances the angular resolution accordingly or allows the measurement at much lower detector-sample distances of, e.g. \sit{2.4}{\m} instead of \sit{8.3}{\m}.
    In general, the sample-detector distance is variable in order to obtain the anticipated electron momentum resolution at sufficiently high count rate. 
    For measurements with lower momentum resolution, this count rate could obviously be even improved significantly.

\section*{Bibliography}
\bibliography{lit}

\providecommand{\newblock}{}
\begin{thebibliography}{10}
\expandafter\ifx\csname url\endcsname\relax
  \def\url#1{{\tt #1}}\fi
\expandafter\ifx\csname urlprefix\endcsname\relax\def\urlprefix{URL }\fi
\providecommand{\eprint}[2][]{\url{#2}}

\bibitem{santhoff1995}
Sauthoff G 1995 {\em Intermetallics\/} (Wiley-VCH)

\bibitem{pfleiderer2009}
Pfleiderer C 2009 {\em Rev. Mod. Phys.\/} {\bf 81}(4) 1551--1624
  \urlprefix\url{https://link.aps.org/doi/10.1103/RevModPhys.81.1551}

\bibitem{lohneysen2007}
L\"ohneysen H~v, Rosch A, Vojta M and W\"olfle P 2007 {\em Rev. Mod. Phys.\/}
  {\bf 79}(3) 1015--1075
  \urlprefix\url{https://link.aps.org/doi/10.1103/RevModPhys.79.1015}

\bibitem{shoenberg2007}
Shoenberg D 2007 {\em Magnetic Oscillations in Metals\/} (Cambridge University
  Press)

\bibitem{cooper1985}
Cooper M~J 1985 {\em Reports on Progress in Physics\/} {\bf 48} 415
  \urlprefix\url{https://dx.doi.org/10.1088/0034-4885/48/4/001}

\bibitem{ketels2021}
Ketels J, Billington D, Dugdale S~B, Leitner M and Hugenschmidt C~P 2021 {\em
  Phys. Rev. B\/} {\bf 104}(7) 075160
  \urlprefix\url{https://link.aps.org/doi/10.1103/PhysRevB.104.075160}

\bibitem{damascelli2003}
Damascelli A, Hussain Z and Shen Z~X 2003 {\em Rev. Mod. Phys.\/} {\bf 75}(2)
  473--541 \urlprefix\url{https://link.aps.org/doi/10.1103/RevModPhys.75.473}

\bibitem{hautojarvi1979}
Hautojärvi P 1979 {\em Positrons in Solids\/} ({\em Topics in Current
  Physics\/} vol~12) (Springer)

\bibitem{ceeh2016}
Ceeh H, Weber J~A, B{\"o}ni P, Leitner M, Benea D, Chioncel L, Ebert H,
  Min{\'a}r J, Vollhardt D and Hugenschmidt C 2016 {\em Scientific reports\/}
  {\bf 6} 20898

\bibitem{weber2015}
Weber J~A, Bauer A, B\"oni P, Ceeh H, Dugdale S~B, Ernsting D, Kreuzpaintner W,
  Leitner M, Pfleiderer C and Hugenschmidt C 2015 {\em Phys. Rev. Lett.\/} {\bf
  115}(20) 206404
  \urlprefix\url{https://link.aps.org/doi/10.1103/PhysRevLett.115.206404}

\bibitem{ketels2021physStat}
Ketels J, Leitner M, Böni P, Hugenschmidt C, Sekania M, James A~D~N, Bonart
  J~A~E, Unglert N and Chioncel L 2022 {\em physica status solidi (b)\/} {\bf
  259} 2100151 (\textit{Preprint}
  \eprint{https://onlinelibrary.wiley.com/doi/pdf/10.1002/pssb.202100151})
  \urlprefix\url{https://onlinelibrary.wiley.com/doi/abs/10.1002/pssb.202100151}

\bibitem{ceeh2013}
Ceeh H, Weber J~A, Leitner M, Böni P and Hugenschmidt C 2013 {\em Review of
  Scientific Instruments\/} {\bf 84} 043905 ISSN 0034-6748 (\textit{Preprint}
  \eprint{https://pubs.aip.org/aip/rsi/article-pdf/doi/10.1063/1.4801454/16149870/043905\_1\_online.pdf})
  \urlprefix\url{https://doi.org/10.1063/1.4801454}

\bibitem{weber2013}
Weber J~A, Böni P, Ceeh H, Leitner M and Hugenschmidt C 2013 {\em Journal of
  Physics: Conference Series\/} {\bf 443} 012092
  \urlprefix\url{https://dx.doi.org/10.1088/1742-6596/443/1/012092}

\bibitem{NIST_gam}
Element/compound/mixture selection
  \urlprefix\url{https://physics.nist.gov/PhysRefData/Xcom/html/xcom1.html}

\bibitem{LYSO_datasheet}
Lyso(ce) scintillator
  \urlprefix\url{https://www.epic-crystal.com/data/upload/20230728/64c32dd1e00bc.pdf}

\bibitem{NaI_datasheet}
Nai(tl) scintillator
  \urlprefix\url{https://www.epic-crystal.com/data/upload/20230728/64c32a167ae2a.pdf}

\bibitem{S13361}
 2022 Mppc ® (multi-pixel photon counter) arrays
  \urlprefix\url{https://www.hamamatsu.com/content/dam/hamamatsu-photonics/sites/documents/99_SALES_LIBRARY/ssd/s13361-3050_series_kapd1054e.pdf}

\bibitem{mppc_kapd9005e}
 2022 Mppc®
  \urlprefix\url{https://www.hamamatsu.com/content/dam/hamamatsu-photonics/sites/documents/99_SALES_LIBRARY/ssd/mppc_kapd9005e.pdf}

\bibitem{PETsys_asic}
Petsys tofpet 2c asic - datasheet (rev 16)
  \urlprefix\url{https://www.petsyselectronics.com/web/website/documentation/TOFPET2%20Downloads/Documentation/PETsys%20TOFPET%202C%20ASIC%20-%20Datasheet%20(rev%2016).pdf}

\bibitem{Schug_2016}
Schug D, Lerche C, Weissler B, Gebhardt P, Goldschmidt B, Wehner J,
  Dueppenbecker P~M, Salomon A, Hallen P, Kiessling F and Schulz V 2016 {\em
  Physics in Medicine and Biology\/} {\bf 61} 2851–2878 ISSN 1361-6560
  \urlprefix\url{http://dx.doi.org/10.1088/0031-9155/61/7/2851}

\bibitem{piatek2014physics}
Piatek S~S, Corporation H and of~Technology N~J~I 2014 Physics and operation of
  an mppc
  \urlprefix\url{https://zeus.phys.uconn.edu/halld/siliconPM/MPPC_physics_of_operation_2014.pdf}

\bibitem{krug2024}
Krug L~M, Chryssos L, Bundesmann J, Dittwald A, Kourkafas G, Denker A and
  Hugenschmidt C 2024 {\em Nuclear Instruments and Methods in Physics Research
  Section B: Beam Interactions with Materials and Atoms\/} {\bf 555} 165488
  ISSN 0168-583X
  \urlprefix\url{https://www.sciencedirect.com/science/article/pii/S0168583X24002581}

\bibitem{kittel}
Kittel C 2004 {\em Introduction to Solid State Physics\/} (John Wiley and Sons
  Inc.) ISBN 9780471415268
  \urlprefix\url{https://www.wiley.com/en-us/Introduction+to+Solid+State+Physics%2C+8th+Edition-p-9780471415268}

\end{thebibliography}

\end{document}